\documentclass{osa-article}

\journal{oe}
\usepackage{cite}
\usepackage[title]{appendix}

\newcommand{\figref}[1]{Fig.~\ref{#1}}
\newcommand{\secref}[1]{Sec.~\ref{sec:#1}}

\renewcommand{\Re}{\operatorname{Re}}
\renewcommand{\Im}{\operatorname{Im}}
\renewcommand{\eqref}[1]{Eq.~(\ref{eq:#1})}
\newcommand{\Eqref}[1]{Equation~(\ref{eq:#1})}
\renewcommand{\vec}[1]{\mathbf{#1}}
\newcommand{\citeasnoun}[1]{Ref.~\citeonline{#1}}

\begin{document}

\title{Approaching the upper limits of the local~density of states via optimized~metallic cavities}

\author{Wenjie Yao$^{1,*}$, Mohammed Benzaouia$^1$, Owen D. Miller$^2$, and Steven G. Johnson$^3$}

\address{$^1$Electrical Engineering and Computer Science Department, Massachusetts Institute of Technology, Cambridge, M.A., 02139, USA\\
$^2$Department of Applied Physics, Yale University, New Haven, C.T., 06520, USA\\
$^3$Department of Mathematics, Massachusetts Institute of Technology, Cambridge, M.A., 02139, USA}

\email{$^*$jayyao@mit.edu} %% email address is required



\begin{abstract}
By computational optimization of air-void cavities in metallic substrates, we show that the local density of states (LDOS) can reach within a factor of $\approx 10$ of recent theoretical upper limits, and within a factor $\approx 4$ for the single-polarization LDOS, demonstrating that the theoretical limits are nearly attainable.   Optimizing the total LDOS results in a spontaneous symmetry breaking where it is preferable to couple to a specific polarization.   Moreover, simple shapes such as optimized cylinders attain nearly the performance of complicated many-parameter optima, suggesting that only one or two key parameters matter in order to approach the theoretical LDOS bounds for metallic resonators.
\end{abstract}

\ocis{(260.3910) Metal optics; (250.5403) Plasmonics; (310.6805) Theory and design.} % REPLACE WITH CORRECT OCIS CODES FOR YOUR ARTICLE, MINIMUM OF TWO; Avoid using the OCIS codes for “General” or “General science” whenever possible.

%%%%%%%%%%%%%%%%%%%%%%%%%%%%%%% Intro %%%%%%%%%%%%%%%%%%%%%%%%%%%%%%%%%%%%
\section{Introduction}

Recently, we obtained theoretical upper bounds~\cite{millerbound} to the (electric) local density of states (LDOS) $\rho(\vec{x},\omega)$, a key figure of merit for light--matter interactions (e.g. spontaneous emission) proportional to the power emitted by a dipole current at a position $\vec{x}$ and frequency $\omega$~\cite{ldos1,ldos2,ldos3,ldos4,OskooiJo13-sources,ldosfree,se1,se2,se3}. For a resonant cavity with quality factor $Q$ (a dimensionless lifetime), LDOS is proportional to the ``Purcell factor'' $Q/V$ where $V$ is a modal volume \cite{purcell1,OskooiJo13-sources}, so LDOS is a measure of light localization in space and time. Our LDOS bounds $\sim |\chi|^2/\Im\chi/ d^3$ (reviewed in~\secref{LDOS}) depend on the material used (described by the $\omega$-dependent susceptibility $\chi = \varepsilon - 1$) and the minimum separation $d$ between the emitter and the material, but are otherwise \emph{independent} of shape, and hence give an upper limit to the localization attainable by \emph{any possible} resonant cavity for $(\chi,d,\omega)$.   However, it is an open question to what extent these bounds are \emph{tight}, i.e. is there any particular cavity design that comes close to the bounds?  Initial investigations of a few simple resonant structures were often orders of magnitude from the upper bounds (except at the surface-plasmon wavelength for a given metal)~\cite{millerbound,raman,bandlimit}.  In this paper, we perform computational optimization of 3D metallic cavities at many wavelengths and find that the bounds are much more nearly attainable than was previously known.

\begin{figure}[tb]
\centering
\includegraphics[width=0.5\linewidth]{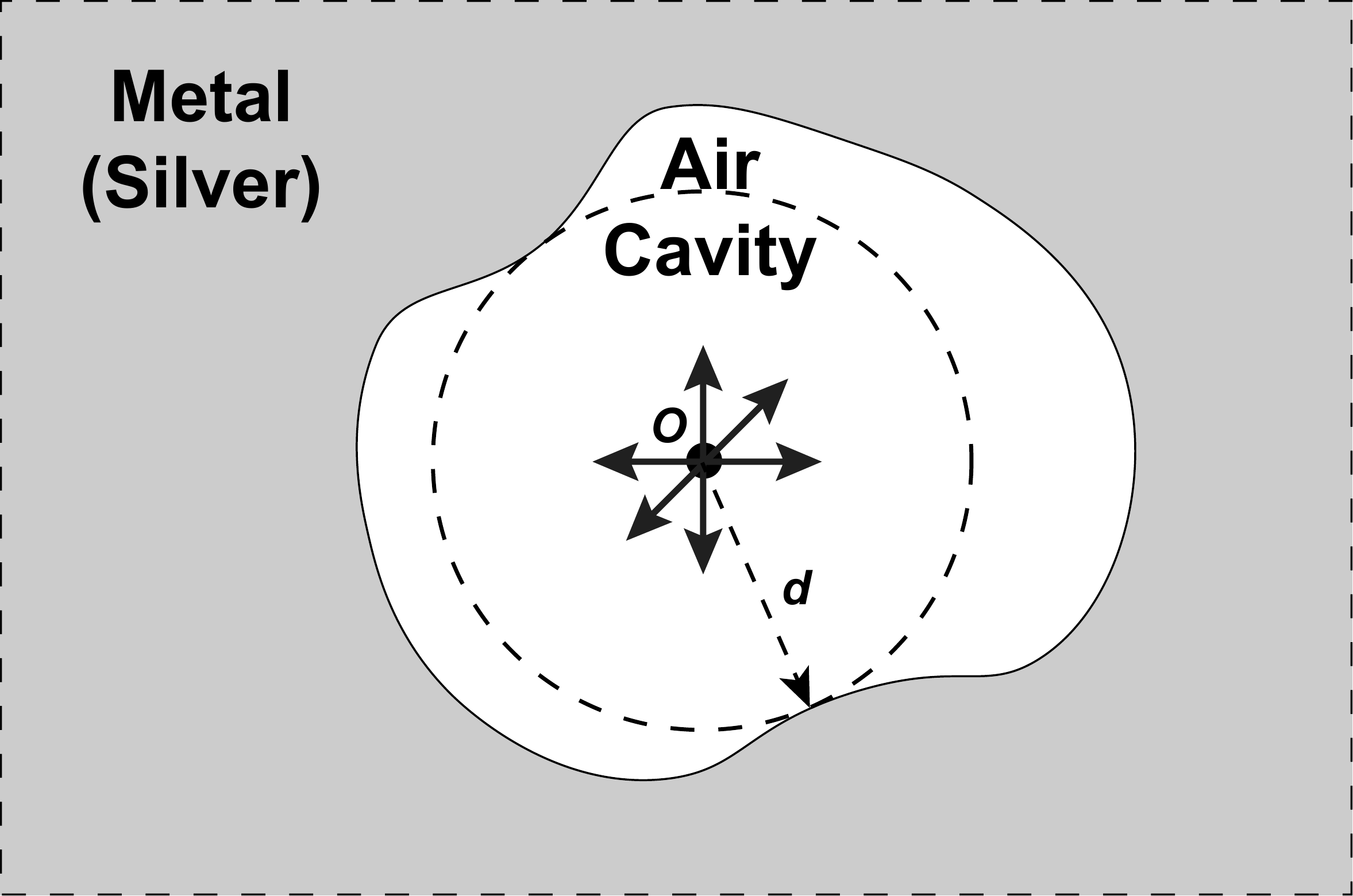}
\caption{Schematic cavity-optimization problem: the shape of an air cavity in a metallic (silver) background is optimized to maximize the LDOS for emitters (dipoles) at the center $o$, constrained for a minimum separation $d$ (the metal lies outside of a sphere of radius $d$). \label{fig1}}
\end{figure}

In particular, we perform many-parameter shape optimization of LDOS for cavities formed by voids in silver (since it has the largest  $|\chi|^2/\Im\chi$, and therefore the largest bounds) at wavelengths $\lambda$ from 400--900 nm and a $d=50$ nm emitter--metal separation, depicted in \figref{fig1}. As described in~\secref{results}, we obtain single-polarization LDOS values within a factor of $\approx 4$ of the theoretical upper bounds, and total (all-polarization) LDOS within a factor of $\approx 10$ of the bounds. Of course, real cavities would have a finite thickness of metal, but our goal is to attain the maximum possible LDOS---we find that a finite-thickness coating has slightly worse performance, but $> 95$\% of the LDOS of the infinite metal is attained by $\approx 100$ nm thickness at $\lambda=500$ nm, and more generally we can theoretically bound~\cite{millerbound} the improvement attainable with \emph{any} additional structure of air voids outside of our cavity. Although our focus is on fundamental upper limits rather than manufacturable cavities, we find that simple shapes (optimized cylinders) are within $\approx 20$\% of the LDOS of optimized many-parameter irregular shapes, analogous to results we obtained previously for optimized scattering and absorption~\cite{millerbound}. (Our optimized cavities are deeply subwavelength along their shortest axes, very different from the non-plasmonic resonant regime where the diameter is much larger than the skin depth so that the walls simply act as mirrors.) Moreover, we find that optimizing for a single emitter polarization (the ``polarized'' LDOS) does nearly as well (within $\approx 10$\%) as optimizing the total LDOS (power summed over all emitter polarizations), reminiscent of earlier results in 2D dielectric cavities where LDOS optimization arbitrarily picked one polarization to enhance~\cite{xiangdong}. We perform the shape optimization using an efficient boundary-element method~\cite{bem,scuff} (which has unknowns only on the metal surface) coupled with adjoint sensitivity analysis~\cite{ad2,ad3} and an optimization algorithm robust to discretization errors~\cite{adam} as described in~\secref{methods}. Although it is possible that even tighter LDOS bounds could be obtained in future results by incorporating additional physical constraints~\cite{add,add1,add2,add3}, we believe that our results show that the existing bounds are already closely related to attainable performance and provide useful guidance for optical cavity design.

In this work, we optimize the \emph{total} electric LDOS, which is the sum of the absorbed and radiated powers from electric dipoles (\secref{LDOS}) for comparison with the upper bounds, and in fact the power is entirely absorbed for an air-void cavity as in \figref{fig1}. However, the same theoretical procedure yielded a bound on the purely \emph{radiative} power that was simply $1/4$ of the total-LDOS bound~\cite{millerbound}, and in fact the two results are closely related.  In general, the addition of low-loss input/output channels can be analyzed via coupled-mode theory as small perturbations to existing cavity designs~\cite{JoannopoulosJo08-book}. As reviewed in Appendix~\ref{append:radiation-TCMT}, given an resonant absorptive cavity, one could modify it to radiate at most $\approx 1/4$ of the original LDOS by slightly perturbing it to add a radiative-escape channel (e.g. a small hole or thinning in the cavity wall) tuned to \emph{match} the absorption-loss rate~\cite{CMT1,CMT2}. (The same channel could also be used to introduce \emph{input} energy, e.g. for pumping an emitter.)  Appendix~\ref{append:shell} shows an example of this: by thinning the cavity walls, we achieve radiated power $\approx 1/4$ of the absorbed power in the purely absorbing cavity.

Many previous authors have computationally optimized the LDOS of cavities (or equivalent quantities such as the Purcell factor $Q/V$), including many-parameter shape or ``topology'' optimization~\cite{opt1,opt2,opt3,opt4,opt5,Wang2018}, but in most cases these works did not compare to the recent upper bounds.  In many cases, these works studied lossless dielectric materials where the bound diverges (though a finite LDOS is obtained for a finite volume~\cite{xiangdong,2dTO} and/or a finite bandwidth~\cite{xiangdong,bandlimit}). Designs specifically for LDOS of metallic resonators that compared to the bounds initially yielded results far below the bounds except for the special case of a planar surface at the surface-plasmon frequency of the material~\cite{millerbound,graphene}, but recent topology optimization in two dimensions came within a factor of 10 of the 2D bound~\cite{2dTO}. Semi-analytical calculations have also been published for resonant modes in spherical metallic voids~\cite{ESM1}, but did not calculate LDOS. Therefore, the opportunity remains for optimized metallic LDOS designs in three dimensions that approach the theoretical upper bounds. To come as close as possible to the bounds, we focus initially on the idealized case of an air void surrounded by metal filling the rest of space, so that there are no radiation losses; later in this paper, we consider the small corrections that arise due to finite metal thickness.

\section{The local density of states (LDOS)}
\label{sec:LDOS}

The (electric) LDOS is equivalent to the total power expended by three orthogonal dipole currents~\cite{ldos1}:
\begin{equation}
\rho=\Im\left[\frac{\epsilon_0}{\pi\omega}\sum_{j=1}^3\hat{\vec{s}}_j\cdot \vec{E}_j(\vec{x}_0)\right],\label{eq:ldoseq}
\end{equation}
where $\epsilon_0$ is the vacuum electric permittivity, $\vec{E}_j$ denotes the field produced by a frequency-$\omega$ unit-dipole source at $\vec{x}_0$ polarized in the $\hat{\vec{s}}_j$ direction, and the sum over $j$ accounts for all three possible dipole orientations. This is equivalent to the average response for \emph{any} dipole orientation~\cite{Wijnands1997}, and is therefore an isotropic figure of merit.   In contrast, we refer to the power $\Im [\frac{\epsilon_0}{\pi\omega} \hat{\vec{s}}_j\cdot \vec{E}_j(\vec{x}_0)]$ expended by only a \emph{single} dipole current as the ``polarized'' LDOS.

From energy-conservation considerations, previous work found an upper bound for LDOS enhancement inside a cavity compared to vacuum electric LDOS ($\rho_0=\omega^2/2\pi^2c^3$~\cite{ldosfree}), given any a material susceptibility $\chi$ and an emitter--material separation $d$ at a frequency $\omega = ck$, to be~\cite{millerbound}:
\begin{equation}
\frac{\rho}{\rho_0} \leq 1+\frac{|\chi(\omega)|^2}{\Im\chi(\omega)}\left[\frac{1}{(kd)^3}+\frac{1}{kd}\right].\label{eq:ldoslimit}
\end{equation}
The details of this bound are reviewed in Appendix~\ref{append:limit}. Two details in \eqref{ldoslimit} require some comment. First, the bounding surface lying between the dipole source and the material for \eqref{ldoslimit} is a sphere of radius $d$ around the source (\figref{fig1}). If the bounding surface is a separating plane, there would be a factor of $1/8$ multiplying the $|\chi|^2/\Im \chi$ term as well as a small modification to the $1/kd$ term~\cite{millerbound}. Second, the separation distance should be small compared to the wavelength, otherwise a third term $\mathrm{O}(kL)$ can have non-negligible contribution to the bound (also discussed in Appendix~\ref{append:sphere}). Finally, the polarized LDOS limit is $1/3$ of the total limit in \eqref{ldoslimit} for the same spherical bounding surface.
%The coefficient in front of the $|\chi|^2/\Im \chi$ term depends on the bounding surface lying between the dipole source and the material.  Here, the material lies outside a sphere of radius $d$ around the source (\figref{fig1}), whereas for a separating plane there is a factor of $1/8$~\cite{millerbound}.   For the polarized LDOS, the $|\chi|^2/\Im \chi$ is multipled by $1/3$ in the bound.

It is important to emphasize that the derivation of~\eqref{ldoslimit} gives a rigorous upper bound to the LDOS, but does not say what structure (if any) achieves the bound.  By actually solving Maxwell's equations for various geometries, we can investigate how closely the bound can be approached (how ``tight'' the bound is).  It is possible that incorporating additional constraints may lead to tighter bounds in the future~\cite{add,add1,add2,add3}, but our results below already show that~\eqref{ldoslimit} is achievable within an order of magnitude.
%%%%%%%%%%%%%%%%%%%%%%%%%%%%%%% Methods %%%%%%%%%%%%%%%%%%%%%%%%%%%%%%%%%%%%
\section{Cavity-Optimization Methods}
\label{sec:methods}
To numerically compute the LDOS inside a metal cavity, we employed a free-software implementation~\cite{scuff} of the boundary element method (BEM)~\cite{bem}. A BEM formulation only involves unknown tangential fields on the metal surface, leading
to modest-size computations for 3D metallic voids (\figref{fig1}). The complex dielectric constant of silver was interpolated from tabulated data~\cite{optcon}.  In addition, we implemented an adjoint method~\cite{ad2,ad3} to rapidly obtain the gradient of the LDOS with respect to the shape parameters described below. As reviewed in Appendix~\ref{append:adjoint}, the gradient of LDOS with respect to all shape parameters simultaneously is obtained by the adjoint method using only two BEM simulations---the original problem and an adjoint problem (the same Maxwell problem with artificial ``adjoint'' sources).   Since the adjoint problem is the same Maxwell/BEM operator, we need only form and factorize the BEM matrix a single time, and the computational cost to solve both the forward and adjoint problems is essentially equivalent to a single simulation.  Validation against a semi-analytical solution for spheres~\cite{ESM1} is discussed in Appendix~\ref{append:sphere}.

In order to parameterize an arbitrary cavity shape numerically, we use a level-set description~\cite{ls1,ls2}, combined with a free-software surface-mesh generator CGAL~\cite{cgal1,cgal2}.  In particular, we describe the radius of the cavity around the source point by a function $R(\theta,\phi)$ (in spherical coordinates), which is expanded below in either spherical harmonics or other polynomials, and equivalently pass a level-set function $\Phi = r - R(\theta,\phi)$ to CGAL (such that $\Phi = 0$ defines the surface).  We considered various parameterizations of the shape function $R$.   The simplest geometries considered were ellipsoids, cylinders, or rectangular boxes, described by two or three parameters.  For many-parameter optimization with a minimum radius (separation) $d$, the function $R$ is expressed as an expansion in some basis functions $S_n(\theta,\phi)$ as:
\begin{equation}
R(\theta,\phi) = d + \left|\sum_{n=0}^N c_nS_n(\theta,\phi)\right|^2.\label{eq:levelsetshape}
\end{equation}
For the basis functions $S_n$, we used either spherical harmonics $Y_{\ell m}(\theta,\phi)$ (for arbitrary asymmetrical ``star-shaped'' cavities) or simple polynomials in $\theta$ (to impose azimuthal symmetry round the $z$ axis and a $z=0$ mirror plane):
\begin{equation}
S_n(\theta,\phi) =\left\{\begin{array}{ll}
\theta^n&\theta\leq\frac{\pi}{2},\\
(\pi-\theta)^n&\theta>\frac{\pi}{2}.\\\end{array}\right.\label{eq:poly}
\end{equation}

The level set is discretized for BEM (by the CGAL software) into a triangular surface mesh.  We used $5\times$ greater resolution for surface points closer to the dipole source (radius $\lesssim 70$~nm), since the singularity of the fields at the source point leads to rapid variations nearby, for around $5000$ triangles overall.  As we deformed the shape during optimization, we first deform the triangles smoothly as long as all angles remained between $30^\circ$ and $120^\circ$, after which point we triggered a re-meshing step.  Unfortunately, re-meshing causes slight discontinuities in the objective function and its derivatives which tend to confuse optimization algorithms expecting completely smooth functions~\cite{noise}.  We tried various optimization algorithms designed to be robust to such ``numerical noise''~\cite{adam,SQP}, and found that the Adam stochastic-optimization algorithm~\cite{adam} seems to work best for our problem.

%%%%%%%%%%%%%%%%%%%%%%%%%%%%%%% Results %%%%%%%%%%%%%%%%%%%%%%%%%%%%%%%%%%%%
\begin{figure}[tb]
\centering
\includegraphics[width=0.99\linewidth]{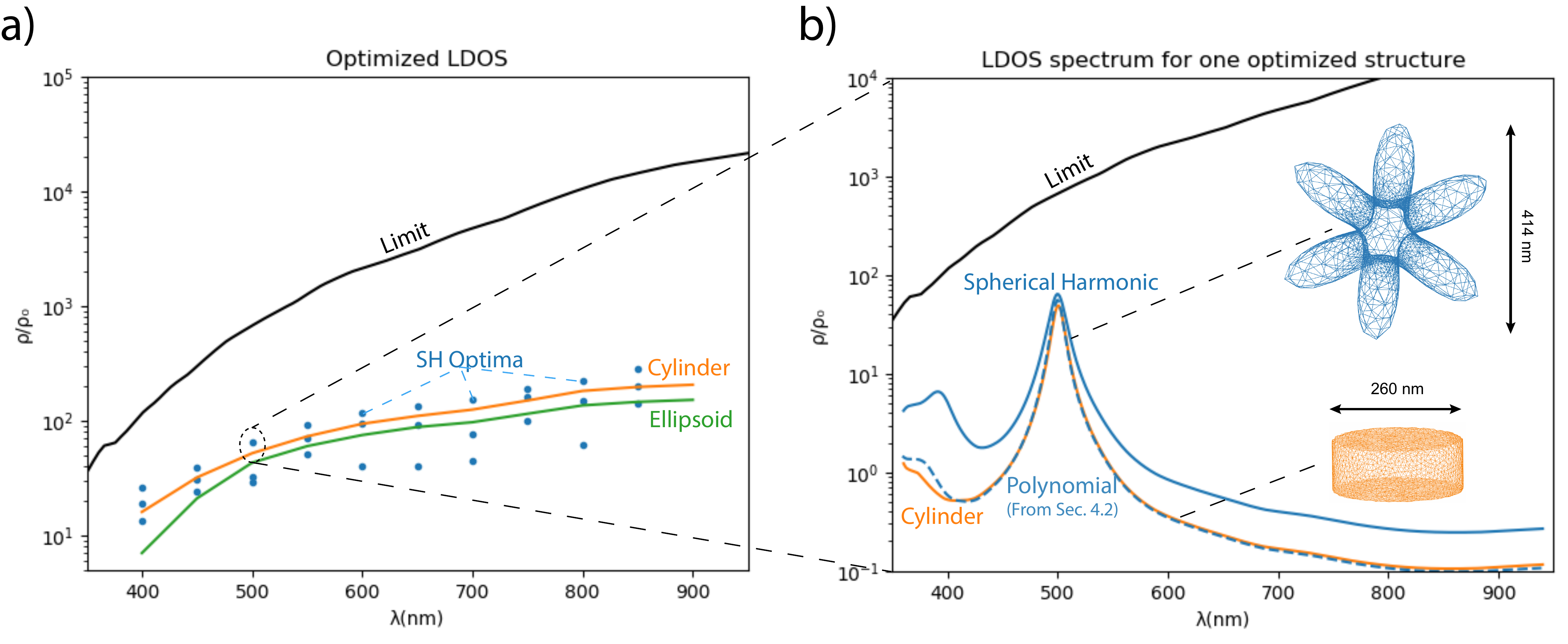}
\caption{(a) Total LDOS optima as a function of the wavelength $\lambda$ for a minimum separation $d=50$~nm, along with the upper bound (black line).
A separately optimized structure is used for each wavelength, either optimized cylinders (orange line) and ellipsoids (green line) or general shape optimization via the optimized spherical-harmonic (SH) surfaces (blue dots) of \eqref{levelsetshape}. Several SH local optima are shown for each $\lambda$, whereas for cylinders and ellipsoids only the global optima are shown. (b) LDOS spectra of the spherical-harmonic (blue) and cylinder (orange) structures optimized for $\lambda=500$~nm, the the shapes (not to scale) inset (see supplementary Visualizations 1 and 2 for 3D views). Also shown is the total-LDOS spectrum of a polynomial shape (dashed blue line) optimized for the polarized LDOS in \secref{polarizedLDOS}, showing that optimizing for a single dipole orientation (polarized LDOS) is nearly equivalent in performance to optimizing for all orientations (total LDOS).  \label{fig2} }
\end{figure}

\section{Cavity-Optimization Results}
\label{sec:results}
\subsection{Total LDOS}
We performed numerical shape optimization of the LDOS for cavities formed by voids in silver~\cite{optcon} at wavelengths $\lambda$ from 400--900~nm, for both simple geometries (cylinder, ellipsoid, and rectangular box) and complex many-parameter shape (spherical harmonics). To obtain a finite optimum LDOS, one must choose a lower bound on the emitter--metal separation distance $d$~\cite{millerbound}. We chose $d=50$~nm so that $kd < 1$ for all optimized wavelengths; this allows us to use \eqref{ldoslimit} as the LDOS upper bound, neglecting additional far-field effects~\cite{millerbound} (see also Appendix~\ref{append:limit}), while a much smaller $d$ was inconvenient to model (due to extremely small feature sizes and even nonlocal effects~\cite{nonlocal1} at such scales).  Note that $d$ also sets a lengthscale for the region of the cavity with maximum LDOS: as long as one shifts the emitter location by $\ll d$ ($\lesssim 10$~nm), the optimized LDOS will be of the similar magnitude, but  other (unoptimized) locations in the cavity will typically have a drastically different LDOS.

The results of the optimized LDOS as a function of the wavelength are displayed in \figref{fig2}a. Note that each wavelength corresponds to a \emph{different} structure optimized for that particular wavelength. If we fix the structure as the one optimized for $\lambda = 500$~nm, the resulting LDOS spectrum is shown in \figref{fig2}b, exhibiting a peak at the optimized wavelength.  For simple shapes (cylinder and ellipsoid), we swept the parameters through a large range and found the global optimum among several local optima.  We also optimized rectangular boxes, but their performance was nearly identical to that of the cylinders (but slightly worse), so they are not shown.   For the 16-parameter (spherical-harmonic) level-set optimization, we performed local optimization for $\sim 10$ random starting points and plotted the best result along with a few other typical local optima.

We found that the optimized LDOS comes within a factor of 10 of the upper bound in the short-wavelength regions ($\lambda<550$ nm), and the optimized cylinders are surprisingly good (within $\approx 20$\% of the many-parameter optima). The optimized cavity geometries at $\lambda=500$~nm  are shown in the inset of \figref{fig2}b. We can see that the optimized many-parameter shape has a three-fold rotational symmetry around one axis; consequently, it has equal polarized LDOS in two directions but $\sim 100\times$ smaller polarized LDOS in the third direction.  The spherical-harmonic basis is unitarily invariant under rotations, so this means that the optimization of the total LDOS (an isotropic figure of merit) exhibits a spontaneous symmetry breaking: it chooses two directions to improve at the expense of the third.  For the optimized cylinder and ellipsoid, the polarized LDOS is only large for \emph{one} polarization (along the cylinder axis, the ``short'' axis).  A similar spontaneous symmetry breaking was observed for optimization of LDOS in two dimensions~\cite{xiangdong}, as well as in saturating the upper bounds for scattering and absorption~\cite{millerbound,sat} (where it was related to quasi-static sum rules constraining polarizability resonances~\cite{sat}).

As discussed below, we found that we could approach the polarized LDOS bound at $\lambda = 500$~nm within a factor of $\sim 3$ for a single dipole orientation; the fact that the total LDOS optimization is worse compared to its bounds ($\approx 3\times{}$ polarized bound) reiterates the conclusion that it is probably not generally possible to maximize the polarized LDOS for all three directions simultaneously.

One possible shape that \emph{will} have same polarized LDOS in all directions is a sphere. As a matter of fact, we found that at each wavelength $\lesssim 600$~nm, there exists a resonant sphere~\cite{ESM1,ESM2} such that the LDOS at the center of the spheres comes within $\approx 20\%$ of the corresponding-$d$ bound (see Appendix~\ref{append:sphere}). However, this resonant radius $d$ is relatively large ($\lambda/4 < d < \lambda/2$, in order to create a resonance at $\lambda$) leading to small LDOS and bound ($\rho / \rho_0 \sim {}$10--100), so saturating such a large-$d$ bound may have limited utility.  Spheres at much smaller $d$ do not exhibit these ``void resonances'' and have much worse LDOS than the asymmetrical shapes in \figref{fig2} for $d=50$~nm.

\begin{figure}[tb]
\centering
\includegraphics[width=0.99\linewidth]{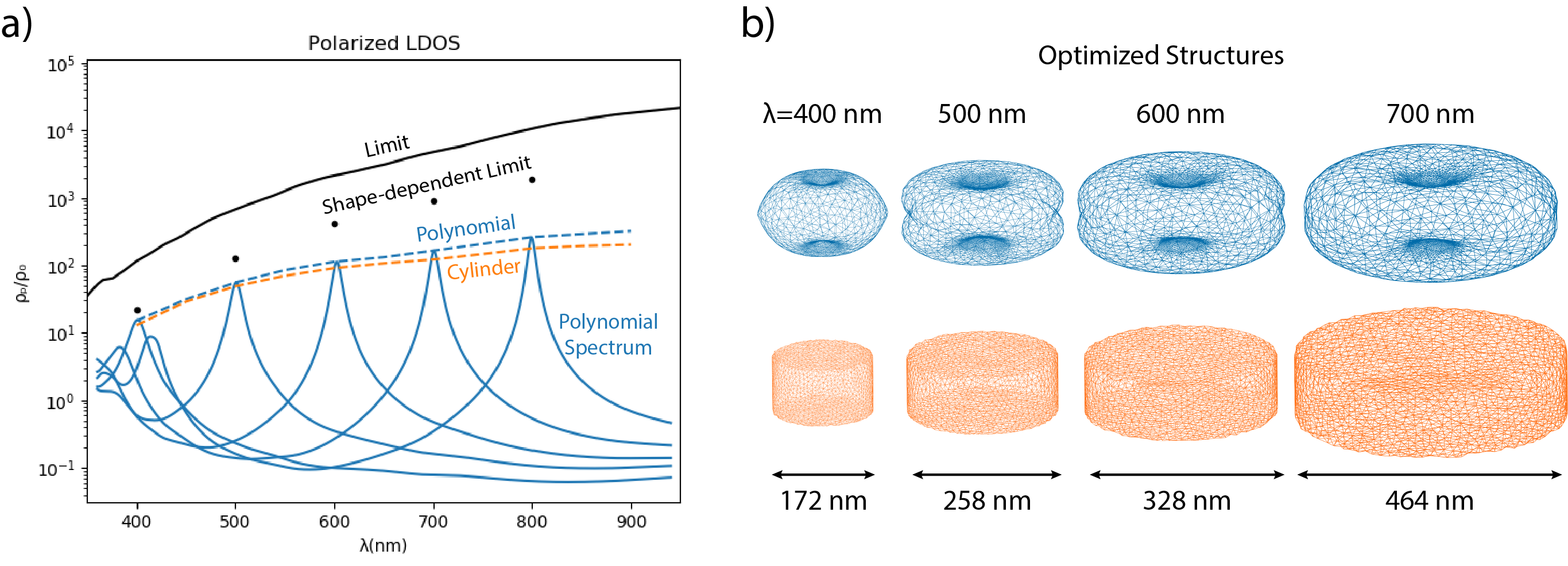}
\caption{(a) Polarized LDOS optima (dashed lines) as a function of the wavelength $\lambda$ at a minimum separation $d=50$ nm, along with the upper bound (black line) and the shape-dependent bounds (black dots). Dashed lines are peak performance of separately optimized structures for each $\lambda$, either cylinders (orange) or the optimized polynomials (blue) of \eqref{poly}. Solid blue lines are spectra of the polarized LDOS for optimized polynomial designs at selected wavelengths $\lambda= 400, 500, 600, 700, 800$~nm, respectively. (b) Optimized polynomial (top) and cylinder (bottom) structures at the wavelength $\lambda= 400, 500, 600, 700$~nm (to scale; see supplementary Visualizations 3--10 for 3D views). \label{fig3} }
\end{figure}

\subsection{Polarized LDOS}
\label{sec:polarizedLDOS}

Since the spontaneous symmetry-breaking in the previous section suggests that optimization favors maximizing LDOS in a single direction, we now consider optimizing the polarized LDOS.  That is, we maximize the power expended by a dipole current with a single orientation (similar to previous work on cavity optimization in dielectric media~\cite{xiangdong,Wang2018,2dTO}).  As above, we performed few-parameter optimization of ellipsoids, cylinders, and rectangular boxes.  For many-parameter optimization, we initially used spherical harmonics but observed that optimizing polarized LDOS naturally leads to structures that are rotationally symmetric around the dipole axis.  To exploit this fact, we switched to simple polynomials in $\theta$ as described in \secref{methods}. Specifically, we first performed a rough scan of degree-2 polynomials to obtain a starting point, then we performed a degree-5 optimization using the adjoint method (degree-10 gave similar results at greater expense). (Gradually increasing the number of degrees of freedom is ``successive refinement,'' a heuristic that has also been used in other work to avoid poor local minima~\cite{suc1,suc2}.) The results are shown in \figref{fig3}. We only plotted the cylinder results (orange dashed line), because the ellipsoid and box results were worse. 

We obtain an optimized LDOS within a factor of about 4 of the polarized-LDOS bound in the short wavelength regions ($\lambda<550$ nm). At a wavelength of 400 nm, the optimized LDOS is only 2.5 times smaller than the bound, which greatly improves upon previous results that often came only within $10^2$--$10^3$ of the bound~\cite{millerbound,graphene,bandlimit}.   One interesting fact is that the optimized polarized LDOS is actually only slightly smaller ($\approx 10\%$) than the optimized total LDOS (blue dashed line in \figref{fig2}b), which
is consistent with the spontaneous symmetry breaking we commented on above: optimizing total LDOS spontaneously chooses one or two directions to optimize at the expense of all others, and hence is often equivalent to optimizing polarized LDOS.  (If an isotropic LDOS is required by an application, one approach is to maximize the minimum of three polarized LDOSes~\cite{xiangdong}.)

The upper bounds can help us to answer another important question: how much additional improvement could be obtained by introducing additional void structures outside of our cavity?  (For example, by giving the cavity walls a finite thickness.)  An upper bound to this improvement is provided by computing a \emph{shape-dependent limit}: we use the same bounding procedure, but evaluate the limit assuming the material lies \emph{outside our optimized shape} rather than outside of a bounding sphere.  This analysis, which is carried out in Appendix~\ref{append:limit}, shows that our optimized polarized LDOS is nearly reaching this shape-dependent limit as shown by the black dots in \figref{fig3}a. Therefore, little further improvement is possible using additional structures outside of the cavity, which justifies optimizing over simple voids in order to probe the bounds.   

We also explicitly studied the effect of a finite thickness for the metallic walls.   To study the cavity thickness effect, we implemented simulations of cylindrical shells using the optimized cylinder at $\lambda=500$ nm taken from Fig. ~\ref{fig2}b. We found that the LDOS increases monotonically with the shell thickness (\figref{Shell} in Appendix~\ref{append:shell}). A shell thickness about 100~nm yields a polarized LDOS within 5\% of the infinite-thickness result, which is not surprising considering that the skin depth~\cite{Jackson,optcon} of silver is $< 33$~nm for $\lambda > 400$~nm.

%%%%%%%%%%%%%%%%%%%%%%%%%%%%%%% Conclusion %%%%%%%%%%%%%%%%%%%%%%%%%%%%%%%%%%%%
\section{Concluding Remarks}
In this work, we obtain LDOS values within a factor of $\approx 10$ of the total LDOS bound and a factor of $\approx 4$ of the polarized LDOS bound in a many-parameter metal-cavity optimization, showing that these upper bounds are  much more nearly attainable than was previously known~\cite{millerbound}.  Unlike previous work on scattering/absorption by small particles~\cite{sat}, our optimized cavities do not appear to be in the quasi-static regime, since their largest axes are $\sim \lambda/2$, even while their smallest axes are deeply subwavelength and exhibit strong plasmonic effects.
It is possible that further improvements could be obtained by a more extensive search of local optima, or by expanding the search to other classes of cavities beyond ``star-shaped'' structures that can be described by a $R(\theta,\phi)$ level set, e.g. via full 3D topology optimization.  We would also like to systematically explore the attainability of the finite-bandwidth bounds from \citeasnoun{bandlimit}, which are useful for lossless dielectrics (where the single-frequency LDOS bound diverges). Conversely, it is possible that incorporating additional constraints, such as considering a more complete form of the optical theorem, may lower the LDOS bounds~\cite{add,add1,add2,add3}.   There has also been recent interest in the magnetic LDOS, corresponding to magnetic-dipole radiation~\cite{Baranov2017}, and we expect that qualitatively similar results (albeit with different optimal shapes) would be obtained for the magnetic LDOS (or some combination of magnetic and electric, although trying to optimize both simultaneously would likely encounter difficulties similar to those for optimizing multiple polarizations).  The magnetic LDOS would merely require one to replace our electric-dipole source with a magnetic-dipole source, along with a similar switch of the Green's function that appears in the derivation of the bounds~\cite{millerbound}.

On a more practical level, a possible next step is to maximize LDOS (or similar figures of merit) for 3D geometries more amenable to fabrication, whereas our goal in the present paper was to probe the fundamental LDOS limits without concern for fabrication.  Fortunately, our results show that relatively simple (constant cross-section) shapes such as cylinders can perform nearly as well as the irregular shapes produced by many-parameter shape optimization, and are relatively insensitive to small details (e.g. curved or flat walls).  This is a hopeful sign for adapting such cavities to nano-manufacturing by lithography or other techniques.  And, although infinite-thickness cavities completely absorb the emitted power, our computation of the radiated power in finite-thickness shells (Appendix~\ref{append:shell}) agrees with the theoretical bound's prediction that the optimal radiated power is $\approx 1/4$ of the total~\cite{millerbound}.

%%%%%%%%%%%%%%%%%%%%%%%%%%%%%%% Appendix %%%%%%%%%%%%%%%%%%%%%%%%%%%%%%%%%%%%
\begin{appendices}
\section{LDOS Limit in Metallic Cavity}
\label{append:limit}
In this appendix, we briefly review the evaluation of the LDOS upper bounds described in Ref.~\citeonline{millerbound}. In particular, the total (electric) LDOS limit can be evaluated as an integral over the entire scattering volume $V$ (the region containing the material $\chi$):
\begin{equation}
\frac{\rho}{\rho_0} \leq 1+\frac{k^3}{4\pi}\frac{|\chi(\omega)|^2}{\Im\chi(\omega)}\int_V\left[\frac{3}{(kr)^6}+\frac{1}{(kr)^4}+\frac{1}{(kr)^2}\right]\mathrm{d}^3r \;,\label{eq:limitint}
\end{equation}
where $\rho_0=\omega^2/2\pi^2c^3$ is the free-space electric LDOS \cite{ldosfree} and $k=\omega/c$ is the wavenumber.

Ostensibly, this limit is dependent on the exact scattering geometry $V$ (leading to a shape-dependent limit). However, \eqref{limitint} is also an upper bound on any scatterer \emph{contained within} $V$~\cite{millerbound}.
In this paper, we are interested in a minimal separation $d$ as depicted in \figref{fig1}, so we take $V$ to be a spherical shell with inner radius $d$ and shell thickness $L$, with $L\rightarrow \infty$ for arbitrary thickness. The integral of \eqref{limitint} can then be evaluated as
\begin{equation}
\frac{\rho}{\rho_0} \leq 1+\frac{|\chi(\omega)|^2}{\Im\chi(\omega)}\left[\frac{1}{(kd)^3}+\frac{1}{kd}+\mathrm{O}(kL)\right]\;,\label{eq:limitO}
\end{equation}
where $\mathrm{O}(kL)$ is a ``Big-O'' asymptotic bound~\cite{bigO}. As discussed in \citeasnoun{millerbound}, the $\mathrm{O}(kL)$ divergence as $L\rightarrow \infty$, which arises from far-field scattering, is unphysical and overly optimistic. The contribution of this term should be limited by the largest
interaction distance over which polarization currents contribute to the LDOS, thus is generally small compared to the first two terms on the right-hand side of \eqref{limitO} and can be neglected at small separation distance $d$. Therefore, the total LDOS limit for a metallic cavity with minimum separation distance $d$ is
\begin{equation}
\frac{\rho}{\rho_0} \leq 1+\frac{|\chi(\omega)|^2}{\Im\chi(\omega)}\left[\frac{1}{(kd)^3}+\frac{1}{kd}\right]\;.\label{eq:limitf}
\end{equation}
Note that this limit, where $V$ is the exterior of a sphere, is about 8 times larger than the limit discussed in \citeasnoun{millerbound} where $V$ was a planar half-space.  In practice, this factor-of-8 improvement may be difficult to realize since optimized cavities will typically have only a small surface area at the minimum separation $d$ (except for the resonant spheres discussed in Appendix~\ref{append:sphere} below).

For the polarized LDOS limit, the integral in \eqref{limitint} (squared Frobenius norm of the homogeneous Green's function~\cite{millerbound}) is replaced with the norm of the dipole polarization vector multiplied by the Green's function norm~\cite{raman}:
\begin{equation}
\frac{\rho_\mathrm{p}}{\rho_0} \leq \frac{1}{3}+\frac{k^3}{8\pi}\frac{|\chi(\omega)|^2}{\Im\chi(\omega)}\int_V\left[a(r)+b(r)|\hat{\vec{p}}\cdot\hat{\vec{r}}|^2\right]\mathrm{d}^3r \;,\label{eq:limitpint}
\end{equation}
where $\hat{\vec{p}}$ is the unit vector in the polarization direction, and $a(r)$ and $b(r)$ are:
\begin{eqnarray}
a(r) &=& \frac{1}{(kr)^6}-\frac{1}{(kr)^4}+\frac{1}{(kr)^2}\\ 
b(r) &=& \frac{3}{(kr)^6}+\frac{5}{(kr)^4}-\frac{1}{(kr)^2} \;.
\end{eqnarray}

Similar to the total LDOS limit analysis, we can use a spherical bounding surface of radius $d$ to derive a general upper bound (also excluding the diverging $\mathrm{O}(kL)$ term):
\begin{equation}
\frac{\rho_\mathrm{p}}{\rho_0} \leq \frac{1}{3}+\frac{|\chi(\omega)|^2}{3\Im\chi(\omega)}\left[\frac{1}{(kd)^3}+\frac{1}{kd}\right]\;,\label{eq:limitp}
\end{equation}
which is exactly $1/3$ of the total LDOS limit. That is, the total LDOS bound is equivalent to assuming that the polarized LDOS bound can be attained for all three polarizations simultaneously, which our results show to be unlikely.

To compute the shape-dependent polarized-LDOS bound (for a given optimized shape) in \secref{polarizedLDOS}, we performed numerical integration of \eqref{limitpint} over spherical angles (with the $r$ integral performed analytically), but excluding the $1/(kr)^2$ radiative term that yields the $\mathrm{O}(kL)$ divergence.

\section{LDOS Gradient from Adjoint Method}
\label{append:adjoint}
The gradient of the LDOS with respect to many shape parameters can be computed by solving Maxwell's equations a \emph{single} additional time (for ``adjoint'' fields) via the adjoint method, a key algorithm for large-scale photonics optimization~\cite{ad2,ad3}.  The specific case of a boundary perturbation is reviewed in \citeasnoun{ophd}, which shows that the variation of an objective function $F$ in response to small shape deformations $\delta R$ (the surface displacement in the normal direction) over the surface $\partial\Omega$ is
\begin{equation}
\delta F = 2\Re\iint_{\partial\Omega} \delta R(\vec{x}^\prime) \left[(1 -\varepsilon) \vec{E}_{\parallel}(\vec{x}^\prime)\cdot\vec{E}^\mathrm{A}_{\parallel}(\vec{x}^\prime) + \left(\frac{1}{\varepsilon}-1\right)\vec{D}_{\perp}(\vec{x}^\prime)\cdot\vec{D}^\mathrm{A}_{\perp}(\vec{x}^\prime)\right]\mathrm{d}S,\label{eq:dF}
\end{equation}
where $\varepsilon$ is the electric permittivity of the metal, $\vec{E}_{\parallel}$ is the surface-parallel electric field, $\vec{D}_{\perp}$ is the surface-parallel displacement field, and the superscript ``$\mathrm{A}$'' denotes the adjoint field excited by an adjoint current source $\vec{J} = \partial F/\partial \vec{E}$.
 
In the case of LDOS, a further simplification arises. The objective function $F=\rho$ at position $\vec{x}_0$ can be expressed as~\cite{ldos1}
\begin{equation}
\rho=\Im\left[\frac{\epsilon_0}{\pi\omega}\sum_{j=1}^3\hat{\vec{s}}_j\cdot \vec{E}_j(\vec{x}_0)\right],\label{eq:ldosexpression}
\end{equation}
where $\vec{E}_j$ denotes the field excited by a unit dipole source at $\vec{x}_0$ polarized in the $\hat{\vec{s}}_j$ direction, and the sum over $j$ accounts for all three possible dipole orientations. We can see from \eqref{ldosexpression} that the LDOS is proportional to the electric field, leading to an adjoint field that is also proportional to the original problem for each orientation $j$,
\begin{equation}
\vec{E}^\mathrm{A}_{j}(\vec{x}) = \Im\left[\frac{\epsilon_0}{\pi\omega}\vec{E}_{j}\right]\;.\label{eq:adf}
\end{equation}
Inserting \eqref{adf} into \eqref{dF} gives us the LDOS gradient (first-order variation) with respect to any shape deformation:
\begin{equation}
\delta \rho =  \frac{\epsilon_0}{\pi\omega}\Im\sum_j\iint_{\partial\Omega} \delta R(\vec{x}^\prime) \left[(1 -\varepsilon) \vec{E}_{j\parallel}(\vec{x}^\prime)^2+ \left(\frac{1}{\varepsilon}-1\right)\vec{D}_{j\perp}(\vec{x}^\prime)^2\right]\mathrm{d}S\;.\label{eq:fderiv}
\end{equation}

\begin{figure}[tb]
\centering
\includegraphics[width=0.5\linewidth]{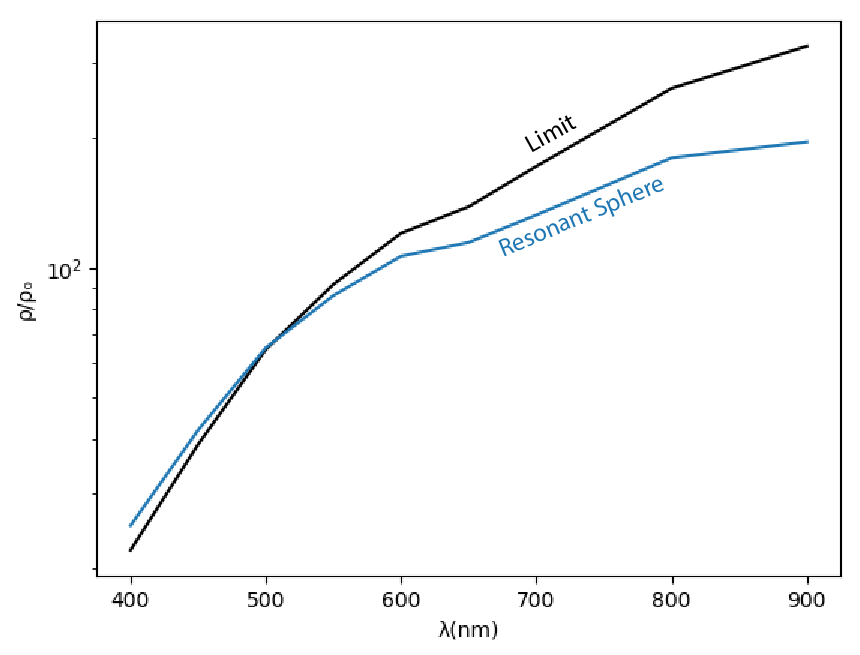}
\caption{LDOS of a resonant air sphere in silver as a function of the wavelength $\lambda$ (blue line), where for each $\lambda$ we choose the smallest radius $a_\mathrm{res}$ for which we couple to a resonant mode at $\lambda$.  The black line is the corresponding upper bound from \eqref{limitf}, setting the minimum separation distance $d = a_\mathrm{res}$.  The LDOS slightly exceeds the bound at small wavelengths where the radius becomes so large that one would need to include the $\mathrm{O}(kL)$ term that we dropped in \eqref{limitO}.\label{LDOSsphere}}
\end{figure}

\section{Resonant Sphere}
\label{append:sphere}
For a void sphere cavity, the resonant electromagnetic surface modes can be analytically obtained by solving the equation~\cite{ESM1,ESM2} (after correcting a typographical error in Ref.~\citenum{ESM1}):
\begin{equation}
\epsilon_\mathrm{m} (\omega) H_\ell(k_\mathrm{m}a)\left[k_\mathrm{d}aJ_\ell(k_\mathrm{d}a)\right]^\prime = \epsilon_\mathrm{d} J_\ell(k_\mathrm{d}a)\left[k_\mathrm{m}aH_\ell(k_\mathrm{m}a)\right]^\prime \;,\label{eq:sphereres}
\end{equation}
where $a$ corresponds to the void radius, $\ell$ is the (integer) index denoting the angular momentum, $k_\mathrm{m}=\sqrt{\epsilon_\mathrm{m}}k$ and $k_\mathrm{d} = \sqrt{\epsilon_\mathrm{d}}k$ are wave vectors in metal and air/vacuum respectively, $J_\ell$ and $H_\ell$ are spherical Bessel and Hankel functions of the first kind, and the prime denotes differentiation with respect to $k_\mathrm{d} a$ or $k_\mathrm{m} a$. Since the excitation source in our LDOS problem is a dipole at the center of the sphere, only an $\ell=1$ mode can be excited. Therefore, for each wavelength $2\pi/k$, there is a minimal resonant sphere: a minimal radius $d=a_\mathrm{res}$ satisfying \eqref{sphereres} for $\ell = 1$. 

Using \eqref{sphereres}, we can directly compute this minimal resonance radius $a_\mathrm{res}$ and then evaluate the corresponding LDOS. Furthermore, since a semi-analytical solution for the resonant modes is known for a void sphere~\cite{ESM1}, the LDOS can also be evaluated analytically via \eqref{ldoseq}. We find that our BEM results are within $2\%$ of the analytical values using a surface mesh of  $\approx 5000$ triangles, validating our numerical solver. The resulting LDOS values are shown in \figref{LDOSsphere}, which shows that the LDOS of the resonant sphere is very close to the theoretical limit (within $\approx 10$\%) for the corresponding minimal separation $d = a_\mathrm{res}$. These strong results verify the limit at least at the resonance combinations of $d$ and $\lambda$. 

Notice that the resonant-sphere LDOS seems to actually slightly exceed the theoretical limit obtained with \eqref{limitf} in short wavelength range. There is no contradiction however: this is simply the effect of the $\mathrm{O}(kL)$ we dropped in \eqref{limitO}.  In particular, the resonant radius here is relatively large compared to the wavelength. For example, at $\lambda=700$~nm we get $a_\mathrm{res}=273$~nm, for which $kd=2.45$ and $1/(kd)^3+1/(kd)=0.47$. As a matter of fact, $1/(kd)^3+1/(kd)<1$ for all resonant spheres, thus the $\mathrm{O}(kL)$ term in \eqref{limitO} will have a non-negligible influence on the bound, causing the actual bound to be slightly higher than \eqref{limitf}.

\begin{figure}[tb]
\centering
\includegraphics[width=0.5\linewidth]{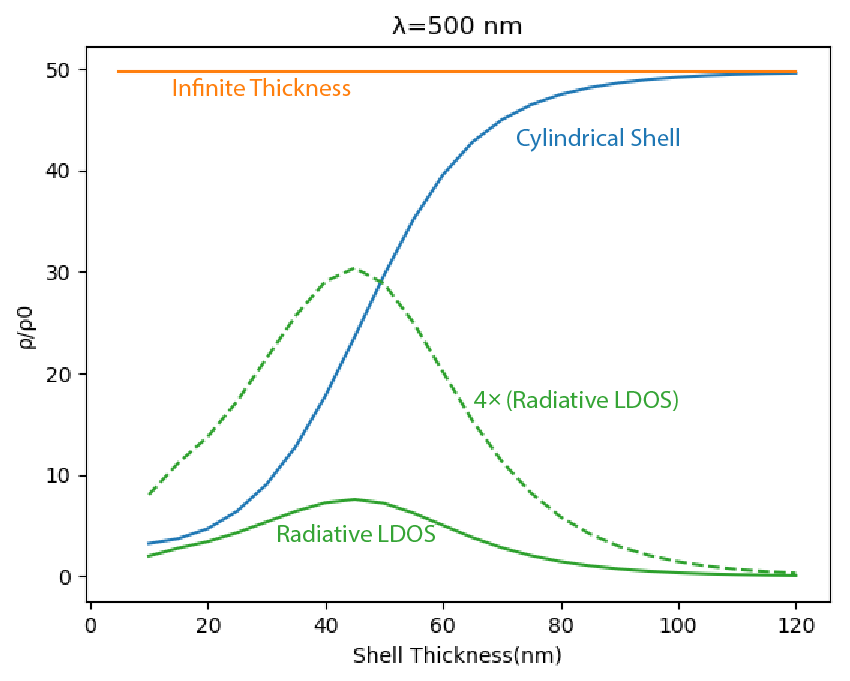}
\caption{ LDOS of optimized cylindrical cavity a function of the wall thickness (blue line) at a wavelength $\lambda=500$~nm and a minimum separation $d=50$~nm, compared to the infinite-thickness LDOS (orange line). Also shown are the \emph{radiative} LDOS $\rho_\mathrm{rad}/\rho_0$ (solid green line: the radiated/non-absorbed power), as well as $4\times \rho_\mathrm{rad}/\rho_0$ (dashed green line) because the theoretical bounds predict that the maximum $\rho_\mathrm{rad}$ is $1/4$ of the total LDOS~\cite{millerbound}.
\label{Shell} }
\end{figure}

\section{Cavity Thickness Effect}
\label{append:shell}
Here, we study the effect of a finite thickness of the metallic walls, replacing the infinite metallic regions of \figref{fig1}.  To do this, we took the optimized cylinder at $\lambda=500$~nm  from Fig. ~\ref{fig2}b and modified the silver walls to have finite thickness with the same inner surface. The LDOS as a function of the shell thickness is shown in \figref{Shell}. We observe that the LDOS increases monotonically with the shell thickness, and that a shell thickness of about 100~nm (about 3.7 times the skin depth) yields a polarized LDOS within 5\% of the infinite-thickness result.

For a finite-thickness shell, some of the expended power (total LDOS) is absorbed and some ``leaks'' through the finite thickness to radiate away, and it is interesting to consider the \emph{radiative} LDOS $\rho_\mathrm{rad}$ defined as the latter radiated power for the same dipole source (green line in \figref{Shell}).  As a function of shell thickness, $\rho_\mathrm{rad}$ exhibits a peak: too thin and the resonance is too weak to enhance LDOS, but too thick and no power escapes to radiate (all power is absorbed).   The theoretical limit for $\rho_\mathrm{rad}/\rho_0 - 1$ is $1/4$ of the limit for the total (absorbed+radiated) LDOS~\cite{millerbound}.  Correspondingly, we plot $4\rho_\mathrm{rad}/\rho_0$ in \figref{Shell} (dashed green line) and see that, at the optimum $\rho_\mathrm{rad}$, the radiative LDOS is approximately $1/4$ of the total ($4\rho_\mathrm{rad} \approx \rho$), agreeing with the prediction of polarization-maximization in Ref.~\citenum{millerbound}.

\section{Coupled-mode theory for radiative LDOS}
\label{append:radiation-TCMT}

Given a purely absorbing resonant cavity such as the ones optimized in this paper, the addition of a radiation-loss channel (e.g. by coupling the cavity to a waveguide, permitting radiation through a small hole in the cavity walls, or simply thinning a portion of the wall as in Appendix~\ref{append:shell}) can be analyzed quantitatively using the technique of temporal coupled-mode theory (TCMT) as long as the lifetime of the cavity remains long enough to be treated as an isolated resonant mode~\cite{JoannopoulosJo08-book,WonjooSuh2004}.  TCMT yields a result identical to that of \citeasnoun{millerbound}: the optimal radiated power is $1/4$ of the original absorbed power.  We describe that straightforward analysis in this appendix, because it helps to connect the results in this paper with applications to radiative cavities.

In particular, consider a purely absorptive cavity with a quality factor~\cite{JoannopoulosJo08-book,WonjooSuh2004} $Q_a \gg 1$, so that the Purcell enhancement of the LDOS for a dipole source is proportional to $Q_a/V$ where $V$ is a corresponding modal volume~\cite{JoannopoulosJo08-book,OskooiJo13-sources}.   Now, suppose that one perturbs the cavity to add a radiative loss channel with a corresponding quality factor $Q_r$.   As long as the radiation loss is low ($Q_r \gg 1$, as is necessary to retain a well-defined resonance), it can be treated as a small perturbation and the effect on $Q_a$ and $V$ can be neglected as higher-order in $1/Q_r$~\cite{JoannopoulosJo08-book}.   With the addition of this channel, the total cavity $Q$ becomes $Q = Q_a Q_r / (Q_a + Q_r)$, corresponding to a total nondimensionalized loss rate of $1/Q = 1/Q_a + 1/Q_r$~\cite{JoannopoulosJo08-book}, so the Purcell enhancement factor is reduced to $Q/V$.  However, only a fraction $Q/Q_r$ of the dipole power goes into radiation (vs. absorption), so the resonant enhancement of the radiated power is proportional to
\begin{equation}
\frac{Q^2}{VQ_r} = \frac{Q_a^2 Q_r}{(Q_a + Q_r)^2 V}. \label{eq:PradQ}
\end{equation}
\Eqref{PradQ} is maximized when $Q_r = Q_a$, i.e. when the absorptive and radiative loss rates are \emph{matched}, similar to results obtained previously~\cite{CMT1,CMT2}.  For $Q_r = Q_a$, \eqref{PradQ} becomes $Q_a/4V$, or exactly $1/4$ of the $Q_a/V$ Purcell enhancement in the purely radiative case.

This result is consistent with the fact that our radiative-LDOS bound is $1/4$ of the total LDOS bound~\cite{millerbound}, but is more far-reaching. It prescribes how any high-$Q$ absorbing cavity can be converted to a radiating cavity with about $1/4$ the radiated power, similar to the results we obtained numerically in Appendix~\ref{append:shell}.

\end{appendices}

%%%%%%%%%%%%%%%%%%%%%%% Acknowledgements

\section*{Funding}
This work was supported in part by the U.S. Army Research Office through the Institute for Soldier Nanotechnologies under award W911NF-13-D-0001, and by the PAPPA program of DARPA MTO under award HR0011-20-90016. O.D.M. was supported by the Air Force Office of Scientific Research under award number FA9550-17-1-0093.

\section*{Disclosures}
The authors declare no conflicts of interest.

%%%%%%%%%%%%%%%%%%%%%%% References %%%%%%%%%%%%%%%%%%%%%%%%%
%\begin{thebibliography}{99}

%\end{thebibliography}
%\bibliographystyle{osajnl}
\bibliography{reference}

\end{document}